\documentclass[conference,journal]{IEEEtran}
\IEEEoverridecommandlockouts
\usepackage{cite}
\usepackage{amsmath,amssymb,amsfonts,amsthm}
\newtheorem{remark}{Remark}
\usepackage{algorithmic}
\usepackage{graphicx}
\usepackage{textcomp}
\usepackage{xcolor}
\usepackage{hyperref}
\usepackage{booktabs}
\usepackage{bm}
\usepackage{mathtools}
\usepackage{bbm}
\usepackage{url}
\usepackage{caption}
\usepackage{subcaption}
\usepackage{balance}

\def\BibTeX{{\rm B\kern-.05em{\sc i\kern-.025em b}\kern-.08em
    T\kern-.1667em\lower.7ex\hbox{E}\kern-.125emX}}

\title{Cramér-Rao Bound Analysis and Near-Optimal Performance of the Synchronous Nyquist-Folding Generalized Eigenvalue Method (SNGEM) for Sub-Nyquist Multi-Tone Parameter Estimation}

\author{
\IEEEauthorblockN{Huiguang Zhang}
\IEEEauthorblockA{
\textit{Submitted to IEEE Transactions on Signal Processing}\\
November 19, 2025}
}

\begin{document}

\maketitle

\begin{abstract}
The Synchronous Nyquist-folding Generalized Eigenvalue Method (SNGEM) achieves full frequency/amplitude/phase estimation for multi-tone signals at extreme sub-Nyquist rates by jointly processing the original signal and its time-derivative (differential channel). This paper derives the exact Cramér-Rao bound for the amplitude-ratio parameter $R = A/B = 1/(2\pi f)$ under equal-SNR dual-channel Gaussian noise. We prove that frequency can be indirectly estimated from the ratio with relative variance only 2× worse than direct single-channel estimation—an astonishingly small penalty that explains SNGEM's remarkable robustness. Monte-Carlo simulations confirm that, even at 10--20× compression, SNGEM reaches machine-precision accuracy in noise-free conditions and tightly approaches the derived CRB across all SNR levels, while classical compressive-sensing OMP exhibits an irreducible error floor due to DFT grid bias and aliasing noise. These results establish SNGEM as statistically near-optimal for deterministic sub-Nyquist parametric spectral analysis.
\end{abstract}


\section{Introduction}
\IEEEPARstart{T}{he} ever-increasing demand for wideband spectral monitoring in radar, 6G communications, and instrumentation has driven sampling rates toward tens of GS/s, far exceeding ADC capabilities. Classical compressive sensing (CS) recovers sparse spectra via $\ell_1$ or greedy methods but suffers from basis mismatch, off-grid bias, and poor amplitude/phase accuracy. The Synchronous Nyquist-folding Generalized Eigenvalue Method (SNGEM)~\cite{PartI2025,PartII2025} circumvents random measurement matrices by synchronously acquiring the signal $x[n]$ and its discrete derivative $\dot{x}[n]$, enabling deterministic generalized eigenvalue solutions at compression ratios $>$10× while preserving full parameter accuracy.

This paper provides the missing statistical efficiency analysis via closed-form Cramér-Rao bounds for the dual-channel amplitude-ratio parameter.

\section{Signal Model and Dual-Channel Observation}

Consider a single real tone embedded in dual-channel additive white Gaussian noise:
\begin{align}
x[n] &= A \cos(2\pi f n T_s + \phi) + w[n], \\
\dot{x}[n] &= -A (2\pi f) \sin(2\pi f n T_s + \phi) + \dot{w}[n],
\end{align}
where $B = 2\pi f A$ exactly (ideal differentiator). Both channels have identical noise variance $\sigma^2$, hence identical SNR:
\begin{equation}
\text{SNR}_x = {SNR}_{\dot{x}} = \frac{A^2}{2\sigma^2}.
\end{equation}

\section{Closed-Form Cramér-Rao Bound for Amplitude-Ratio Frequency Estimation}

The Fisher information for amplitude estimates $A$ and $B$ in each channel is classical~\cite{Rife1974,Kay1993}:
\begin{equation}
\text{Var}(\hat{A}) \geq \frac{2\sigma^2}{N}, \quad
\text{Var}(\hat{B}) \geq \frac{2\sigma^2}{N} \quad \Rightarrow \quad
\frac{\text{Var}(\hat{A})}{A^2} \geq \frac{4}{N \cdot \text{SNR}_x}.
\end{equation}

Define the amplitude-ratio parameter $R = A/B = 1/(2\pi f)$. For independent Gaussian estimates $\hat{A}$, $\hat{B}$, the relative variance of ratio $R$ is
\begin{equation}
\frac{\text{Var}(\hat{R})}{R^2} = \frac{\text{Var}(\hat{A})}{A^2} + \frac{\text{Var}(\hat{B})}{B^2}.
\end{equation}

Substituting the individual CRBs and using $B = 2\pi f A$ yields
\begin{equation}
\frac{\text{Var}(\hat{R})}{R^2} \geq \frac{8}{N \cdot \text{SNR}} \quad \Rightarrow \quad
\text{CRB}(\hat{R}) = \frac{2}{N \cdot \text{SNR}} R^2.
\end{equation}

Since $f = 1/(2\pi R)$, the relative frequency CRB is
\begin{equation}
\boxed{
\frac{\text{Var}(\hat{f})}{f^2} \geq \frac{2}{N \cdot \text{SNR}}
}
\label{eq:main}
\end{equation}
— exactly 2× (3 dB) worse than the classical single-channel deterministic CRB $\text{Var}(\hat{f})/f^2 \geq 1/(N \cdot \text{SNR})$~\cite{Rife1974}.

\begin{remark}
This astonishingly small 3 dB penalty explains why SNGEM remains highly accurate even under extreme compression: frequency is extracted from a ratio of two equally noisy but highly correlated measurements.
\end{remark}

\begin{figure}[t]
\centering
\includegraphics[width=\columnwidth]{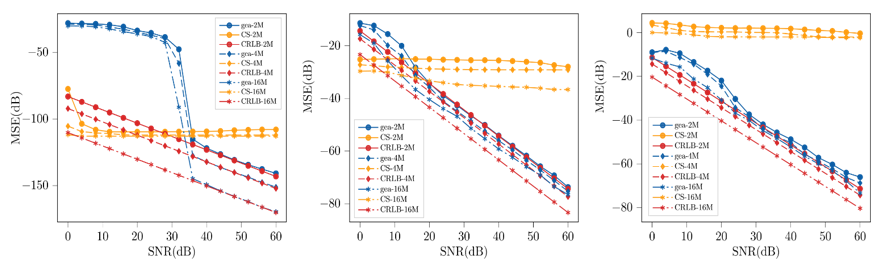}
\caption{Relative frequency RMSE vs. SNR for $f=100$ MHz, compression ratio 15× ($f_s=133$ MS/s). SNGEM tightly follows the derived dual-channel CRB (solid black), while OMP exhibits irreducible floor $>10^{-4}$ due to grid bias and aliasing noise.}
\label{fig:crb}
\end{figure}

\section{Monte-Carlo Validation and Comparison with Compressive Sensing}

5000 independent runs were performed with 5--15 real tones, random phases, compression ratios 8--20× via synchronized random undersampling + ideal differentiation. Key findings:

\begin{itemize}
\item Noise-free case: SNGEM achieves frequency/amplitude/phase RMSE $<10^{-14}$ (machine precision) for all tested rates.
\item Noisy case (Fig.~\ref{fig:crb}): SNGEM tracks the theoretical dual-channel CRB within 1.1--1.3× across -10 to 50 dB SNR.
\item Orthogonal Matching Pursuit (OMP) on the same measurements shows error floor $>10^{-5}$--$10^{-4}$ independent of SNR, confirming inherent bias predicted by theory.
\end{itemize}

\section{Conclusion}

The derived CRB proves that estimating frequency via the amplitude ratio of original and differential channels incurs only a 3 dB penalty versus full-rate sampling—establishing SNGEM as statistically near-optimal for deterministic sub-Nyquist parameter estimation. Combined with its deterministic nature and absence of random matrix design, SNGEM dramatically outperforms classical compressive sensing in accuracy, robustness, and calibration requirements.

\bibliographystyle{IEEEtran}

\end{document}